\documentclass[preprint,12pt]{elsarticle}

\usepackage{amssymb}
\usepackage{amsmath}
\usepackage{algorithm}
\usepackage{algpseudocode}
\usepackage{braket}

\usepackage{lineno}

\journal{Nuclear Physics B}

\begin{document}

\begin{frontmatter}



\title{Single-shot quantum neural networks with amplitude estimation}


\author[1,*]{Jaemin Seo} 

\affiliation[1]{organization={Department of Physics, Chung-Ang University},
            city={Seoul 06974},
            country={Republic of Korea}}
\affiliation[*]{Corresponding author's e-mail: jseo@cau.ac.kr}

\begin{abstract}
Quantum neural networks (QNNs) suffer from a fundamental sampling bottleneck since quantum measurements are probabilistic, requiring many circuit executions to estimate outputs with sufficient accuracy. Conventional Monte-Carlo (MC) inference exhibits an $\mathcal{O}(1/\sqrt{N})$ sampling error, rendering QNN inference and training costly on near-term quantum hardware, especially where each shot requires expensive qubit generation. This work introduces a ``single-shot'' QNN framework by integrating quantum amplitude estimation (AE) into the readout stage. By embedding a trained QNN as a state-preparation oracle within AE, outputs are estimated through coherent interference rather than repeated sampling. We demonstrate that AE-based QNN inference achieves an $\mathcal{O}(1/N)$ error even with a single shot. We further analyze noise robustness and training feasibility, showing that AE can be a powerful primitive for overcoming the sampling overhead of QNNs. This highlights that when the model itself is quantum, quantum algorithms can enhance the computation efficiency.

\end{abstract}


\end{frontmatter}



\section{Introduction}\label{sec1}

The rapid development of quantum computing has opened new ways for solving computational problems that are intractable for classical computers. Using intrinsically quantum mechanical phenomena such as superposition and entanglement, quantum devices promise algorithmic advantages in areas including optimization, simulation, and probabilistic inference \cite{shor_1999_qc, prl_2017_grover_algo, prl_2009_hhl_algo, SCHALKERS2024112816}. In parallel, the recent success of machine learning has stimulated intense interest in combining quantum computation with learning paradigms, giving rise to the rapidly growing field of quantum machine learning (QML) \cite{Biamonte2017_qml, Cerezo2022_qml, seo2026efficientquantummachinelearning}. Within this context, quantum neural networks (QNNs) have emerged as a prominent class of variational quantum models, aiming to extend the expressive power of classical neural networks into the quantum domain \cite{Benedetti_2019_pqc, Hubregtsen2021_pqc, ieee2023_pqc}.

A QNN typically consists of a parametrized quantum circuit in which classical input data are encoded into qubit states through unitary rotations, followed by layers of trainable quantum gates. After the circuit evolution, information is extracted from the final quantum state via measurement, producing an output that can be interpreted as a probability or expectation value. QNNs offer several attractive features, including compact representations in high-dimensional Hilbert spaces, natural compatibility with quantum data, and the potential for favorable scaling in specific tasks \cite{Wen2024_CP_qnn_expressivity, Panadero2024_SR_qnn_expressivity}. As such, they have been investigated in a wide range of applications, from classification and regression \cite{seo_ieee_2025, seo_ieee_2026} to physics-informed modeling \cite{10.1063/5.0226232_qpinn, qpinn_PRA2026} and surrogate representations of complex quantum systems \cite{doi:10.34133/research.0134_qnn_for_quantum}.

Despite these advantages, QNNs face a fundamental difficulty that has no direct analogue in classical neural networks. In a classical model, a single forward pass deterministically produces an output $y$ for a given input $x$. However, due to the intrinsic uncertainty of quantum measurements, the output of a QNN is not a deterministic value but a random variable whose statistics are governed by the Born rule. Consequently, the desired output must be inferred from repeated executions of the same circuit with the same input, each followed by a measurement, and the final prediction is obtained as an empirical expectation value.

This inference procedure is effectively a Monte Carlo (MC) sampling process. For a binary measurement outcome, estimating an output probability from $N$ repeated shots results in a standard error that scales $\mathcal{O}(1/\sqrt{N})$. This sampling error imposes a significant computational burden; achieving even modest accuracy $\sim 1\%$ typically requires tens of thousands of circuit executions. On superconducting \cite{Clarke2008_superconducting} or trapped-ion platforms \cite{Pino2021_trapped_ion}, such large shot counts induce substantial time overhead due to repeated initialization, measurement, and control cycles. The situation is even more difficult for photonic quantum platforms \cite{Alexander2025_photonic_quantum}, where each shot requires the physical generation of new qubits (photons). In many photonic implementations, producing a single effective photon can demand hundreds to thousands of high-power laser pulses, making multi-shot MC-based inference prohibitively expensive. The limitations of multi-shot inference extend beyond hardware considerations. In certain applications, repeated measurements are intrinsically restricted. Examples include inference on invasive or fragile biological samples, experiments involving high-cost optical setups, or scenarios in which one aims to predict extremely rare events with very small probabilities. In such cases, the requirement of large numbers of identical circuit executions makes conventional MC-based readout unsuitable, regardless of the underlying quantum hardware.

From an algorithmic perspective, however, this sampling bottleneck is not fundamental. For probability estimation problems, quantum amplitude estimation (AE) provides a principled mechanism to reduce the required resources quadratically, from $\mathcal{O}(N)$ to $\mathcal{O}(\sqrt{N})$, compared to classical MC sampling. Importantly, this reduction does not stem from repeated measurements but from coherent quantum interference enabled by Grover-type iterations. While the $\sqrt{N}$ scaling reflects the number of controlled Grover operations applied within the circuit, the number of measurement shots required for readout can be reduced to a ``single shot'' or a few shots while maintaining high estimation accuracy.

In this work, we introduce a framework that integrates quantum amplitude estimation with quantum neural networks to dramatically reduce the resources and shots required for QNN inference. By embedding a fixed QNN within an amplitude estimation protocol, we demonstrate that output probabilities can be inferred with accuracy comparable to conventional MC-based readout while using orders of magnitude fewer measurement shots. Remarkably, accurate inference is achievable even in the single-shot limit. This capability is particularly relevant for photonic quantum platforms, where qubit generation constitutes a dominant cost, and suggests a viable pathway toward practical QML inference on hardware where standard sampling-based approaches are infeasible. More broadly, our results highlight a key principle in quantum computation. When the model itself is quantum, quantum algorithms can offer decisive advantages not only in training or optimization but also in the readout and inference stages. By explicitly exploiting quantum algorithms for probabilistic estimation, we show that the apparent sampling overhead of QNNs is not an inherent limitation but can be overcome through algorithmic design. This study therefore provides both a computational and conceptual contribution, illustrating how quantum algorithms can fundamentally reshape the cost structure of QML models.

\section{Background}\label{sec2}

\subsection{Quantum neural networks}\label{sec2.1}

QNNs are commonly formulated using parameterized quantum circuits (PQCs) \cite{Benedetti_2019_pqc, Hubregtsen2021_pqc, ieee2023_pqc}, which serve as variational models operating on quantum states. A PQC consists of a sequence of unitary operations whose structure is fixed, while a set of continuous parameters is optimized through classical computational feedback. Given their analogy to classical neural networks, where parameters are updated to minimize a loss function, PQCs provide a natural foundation for constructing QNNs.

\begin{figure}[t!]
\centering
\includegraphics[width=\linewidth]{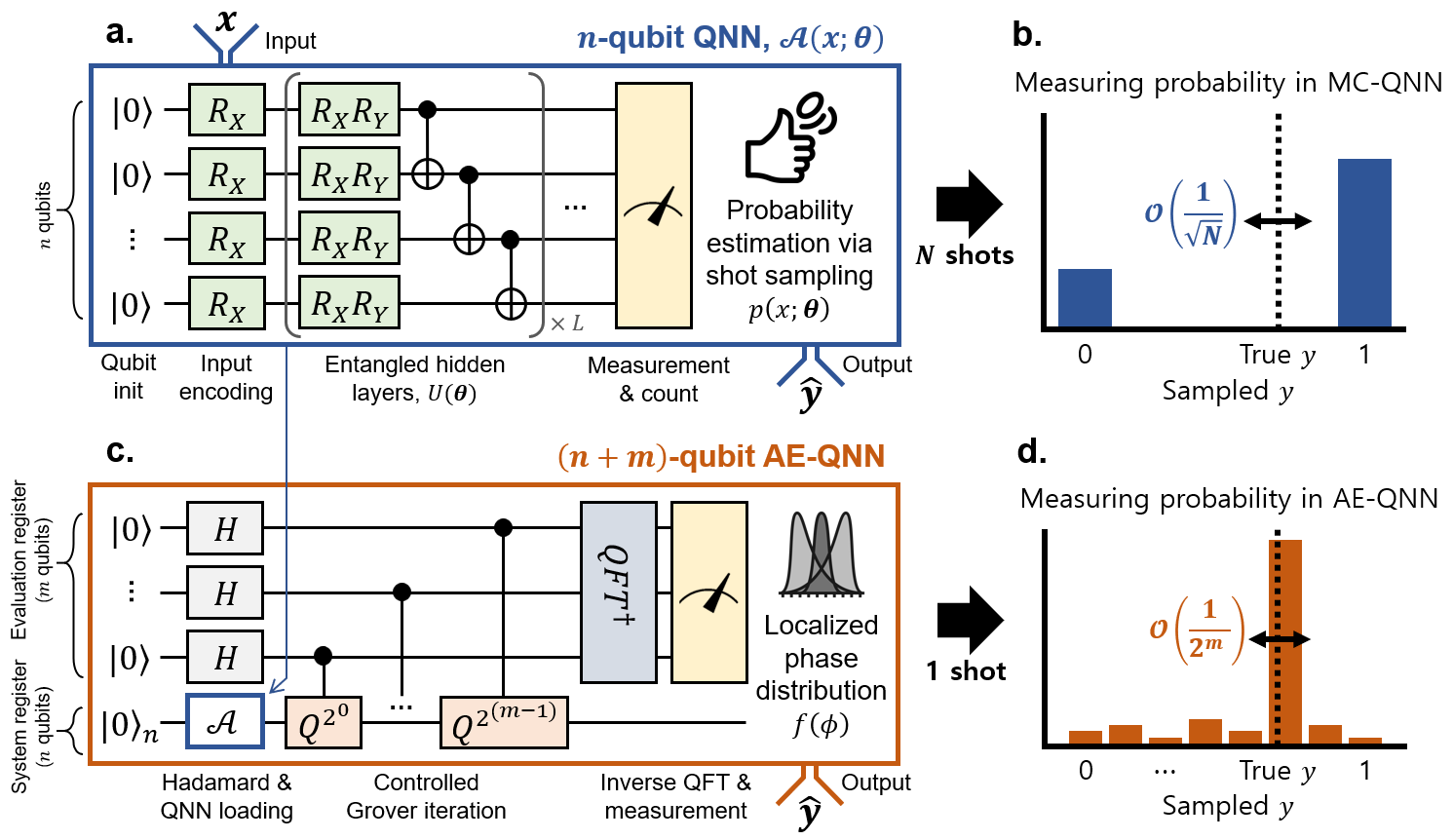}
\caption{Illustration of the architecture of QNN and our proposed inference framework. \textbf{a}. A conventional QNN inference framework using Monte Carlo sampling. \textbf{b}. Measurement probability of the conventional sampling, which gives the error scaling $\mathcal{O}(1/\sqrt{N})$ with measurement shots $N$. \textbf{c}. Our proposed QNN inference framework using a quantum amplitude estimation protocol. \textbf{d}. Measurement probability of the proposed sampling, which gives the error scaling $\mathcal{O}(1/2^m)$ with the evaluation register qubits $m$, even with a single measurement shot.}\label{fig1}
\end{figure}

Consider an $n$-qubit quantum register initialized in a reference state $\ket{0}^{\otimes n}$, as shown in Figure \ref{fig1}a. For a given classical input $x$, data are encoded into the quantum state through an input-dependent unitary operation $U_\text{enc}(x)$. This is followed by a parameterized unitary operator 

\begin{equation}
    U(\boldsymbol{\theta}) = \prod^L_{l=1} U_l(\boldsymbol{\theta}_l),
\label{eq_unitary_transform} \\
\end{equation}

\noindent where $\boldsymbol{\theta}=\{\theta_1, ..., \theta_P\}$ denotes the trainable parameters of the QNN. The overall state preparation can thus be written as 

\begin{equation}
    \ket{\psi(x;\boldsymbol{\theta})} = U(\boldsymbol{\theta}) U_{\text{enc}}(x) \ket{0}^{\otimes n}.
\label{eq_quantum_state} \\
\end{equation}

The circuit $U=\{U_l\}$ is typically composed of alternating layers of single-qubit rotations and entangling gates, ensuring sufficient expressivity. The role of learning is to adjust $\boldsymbol{\theta}$ such that the quantum state $\ket{\psi(x;\boldsymbol{\theta})}$ encodes task-relevant information about the input $x$.

Unlike classical neural networks, the output of a QNN cannot be directly read from the final state vector. Instead, information is extracted via quantum measurement. Let $O$ be a Hermitian observable acting on the system, or equivalently a projector corresponding to a ``good'' subspace of interest ($\ket{\psi_1}$), such as $\ket{1}^{\otimes n}$. The QNN output is defined as the expectation value

\begin{equation}
    p(x;\boldsymbol{\theta}) = \bra{\psi(x;\boldsymbol{\theta})} O \ket{\psi(x;\boldsymbol{\theta})},
\label{eq_measurement} \\
\end{equation}

\noindent which can be interpreted as a probability or regression output depending on the application.

In practice, this expectation value cannot be accessed in a single execution of the circuit. Instead, the circuit must be executed repeatedly, and the observable must be measured multiple times. For a binary measurement outcome, each circuit execution produces a Bernoulli random variable, and the estimator $\hat p$ obtained from $N$ repeated shots satisfies

\begin{equation}
    \text{Var}(\hat p) = \frac{p(1-p)}{N},
\label{eq_sample_noise} \\
\end{equation}

\noindent implying a standard error that scales as $\mathcal{O}(1/\sqrt{N})$, as illustrated in Figure \ref{fig1}b. This Monte Carlo (MC) sampling noise is intrinsic to quantum measurement and constitutes a fundamental overhead in QNN inference.

Training a QNN proceeds by minimizing a classical loss function $\mathcal{L}$, defined in terms of the measured outputs $p(x;\boldsymbol{\theta})$ and target values. Gradient-based optimization methods are commonly employed, requiring the evaluation of partial derivatives $\partial \mathcal{L}/\partial\theta_i$, or equivalently $\partial p/\partial\theta_i$. For PQCs, these derivatives can be computed exactly using the parameter-shift rule \cite{Mitarai_PRA2018_param_shift, Schuld_PRA2019_param_shift}, which for a single parameter $\theta_i$ takes the form

\begin{equation}
    \frac{\partial p}{\partial \theta_i} = \frac{1}{2}\left[p(x;\boldsymbol{\theta}_i^+) - p(x;\boldsymbol{\theta}_i^-) \right],
\label{eq_param_shift} \\
\end{equation}

\noindent where $\boldsymbol{\theta}_i^\pm = \boldsymbol{\theta} \pm (\pi/2) \textbf{e}_i$, and $\textbf{e}_i$ is the unit vector in parameter space.

As a result, computing the gradient with respect to $P$ parameters requires at least $P+1$ distinct circuit evaluations per training step; one for the forward pass and $P$ for each parameter shift. When each expectation value is itself estimated using $N$ shots, the total number of circuit executions per optimization step scales as $(P+1)N$. This quickly leads to a substantial sampling cost, particularly for QNNs with many parameters or for datasets requiring repeated evaluations.

Consequently, QNN training is significantly more resource-intensive than inference. While inference requires repeated shots only to estimate a single expectation value, training involves repeated measurements across many parameter-shifted circuits. This disparity is especially problematic for platforms where shot execution is expensive.

\subsection{Quantum amplitude estimation}\label{sec2.2}

Quantum amplitude estimation (AE) is a quantum algorithm designed to estimate the probability amplitude associated with a target subspace more efficiently than classical MC sampling. Its foundation lies in Grover's algorithm \cite{prl_2017_grover_algo} and the principle of amplitude amplification \cite{Grinko2021_iqae}.

Consider a unitary operator (or an oracle) $\mathcal{A}$ that prepares a quantum state

\begin{equation}
    \mathcal{A} \ket{0}^{\otimes n} = \sqrt{1-a} \ket{\psi_0} + \sqrt{a} \ket{\psi_1},
\label{eq_ae_operator} \\
\end{equation}

\noindent where $\ket{\psi_1}$ spans the ``good'' subspace ($\ket{\psi_0}$ is the other) and $a \in [0, 1]$ is the target probability to be estimated. Classical MC estimation of $a$ requires $\mathcal{O}(1/\epsilon^2)$ samples to achieve an additive error $\epsilon$.

Grover's amplitude amplification constructs a unitary operator 

\begin{equation}
    Q = -\mathcal{A} S_0 \mathcal{A}^{\dagger} S_\chi,
\label{eq_grover_aa} \\
\end{equation}

\noindent where $S_0$ and $S_\chi$ are selective phase reflections about the initial state and the good subspace, respectively. Repeated application of $Q$ performs a rotation in the two-dimensional subspace spanned by $\left\{ \ket{\psi_0}, \ket{\psi_1} \right\}$, coherently amplifying the amplitude of the good state. Canonical AE combines amplitude amplification with quantum phase estimation \cite{PhysRevLett.102.040403_phase_estimation} enabled by this rotation structure. Writing $a=\sin^2{\phi}$, the Grover operator $Q$ has eigenvalues $e^{\pm 2i\phi}$. By applying quantum phase estimation to $Q$, one can estimate the phase $\phi$ and thereby recover $a$.

In its standard formation, AE employs an evaluation register of $m$ qubits and performs a sequence of controlled Grover operations $Q^{2^k}$ for $k=0, \dots m-1$, followed by an inverse quantum Fourier transform (QFT). Measuring the evaluation register yields an $m$-bit estimate of the phase, from which the amplitude is reconstructed as

\begin{equation}
    \hat a = \sin^2(\pi z/2^m),
\label{eq_amplitude_recover} \\
\end{equation}

\noindent where $z \in \{0,1, \dots, 2^{m-1}-1, 2^{m-1} \}$ denotes the measurement outcome index. The resulting estimation error scales as $\mathcal{O}(1/2^m)$, corresponding to a quadratic improvement over classical MC methods in terms of the number of Grover queries $N_\text{query}=2^m$.

A crucial distinction is that this quadratic speedup is achieved through coherent quantum evolution rather than repeated measurements. While the canonical AE circuit contains $\mathcal{O}(2^m)$ Grover iterations, the number of measurement shots required to extract the estimate can be as low as a single shot or a few shots to suppress readout noise.

\section{Quantum neural networks with amplitude estimation}\label{sec3}

\subsection{Integrating amplitude estimation with quantum neural networks}\label{sec3.1}

The sampling problem of the QNN outputs described in Section \ref{sec2.1} naturally fits the AE framework shown in Section \ref{sec2.2}. As discussed before, a QNN output defined as an expectation value, $p(x;\boldsymbol{\theta}) = \bra{\psi(x;\boldsymbol{\theta})} O \ket{\psi(x;\boldsymbol{\theta})}$, corresponds precisely to the amplitude of a ``good'' subspace. In conventional implementations, this quantity is estimated via repeated measurements and MC averaging, incurring a sampling error of $\mathcal{O}(1/\sqrt{N})$.

By embedding the QNN state-preparation unitary $\mathcal{A}(x; \boldsymbol{\theta})$ into an AE protocol, the same output probability can be estimated with quadratically fewer effective iterations. Structurally, this integration treats the entire QNN as a black-box state-preparation oracle within the amplitude estimation circuit, as shown in Figure \ref{fig1}c. From the perspective of thre readout stage, the QNN is no longer repeatedly sampled but instead interrogated coherently through Grover-type iterations. The structural decoupling is particularly significant for QNNs. Unlike classical neural networks, the output of a QNN need not be directly mapped to a classical wire at the end of the circuit, as shown in Figure \ref{fig1}c. The measurement can be deferred and embedded within a larger quantum algorithm, such as AE, without altering the definition of the model.

From the training point of view, the training of classical neural networks relies on backpropagation, which requires that the output be connected to all parameters through a differentiable computational graph. The output value must therefore be explicitly computed and stored as a classical variable in order to propagate gradients backward through the network. QNNs differ fundamentally in this regard. Gradients with respect to circuit parameters can be obtained using the parameter-shift rule, which requires only forward evaluations of the quantum circuit at shifted parameter values. As a result, the output observable does not need to be explicitly extracted as a classical quantity at the end of the circuit. Instead, it suffices that the measurement statistics encode the desired expectation value. This distinction allows the QNN output to remain embedded within a quantum subroutine, as illustrated schematically in Figure \ref{fig1}c, without violating the requirements of gradient-based optimization. Consequently, integrating AE into a QNN does not obstruct training from an algorithmic standpoint, even though the circuit structure of the readout stage differs substantially from classical inference.

The combination of QNNs with amplitude estimation highlights a broader principle. When the model itself is quantum, quantum algorithms can be leveraged not only for state preparation and evolution but also for inference. In settings where measurement shots are expensive, such as photonic platforms, invasive experiments, or rare-event estimation, the ability to replace repeated MC sampling with coherent AE offers a compelling pathway toward practical quantum machine learning. In the following sections, we build on this theoretical background to demonstrate how AE can be systematically combined with QNNs to achieve shot-efficient inference, and we analyze the resulting trade-offs between circuit depth, noise sensitivity, and estimation accuracy.

\subsection{Experimental setting}\label{sec3.2}

\begin{algorithm}[t]
\caption{Monte Carlo inference for QNN output (MC-QNN)}
\label{alg1}
\begin{algorithmic}[1]

\Require
QNN state-preparation circuit $\mathcal{A}(x;\boldsymbol{\theta})$ acting on $n$ qubits,
Projector $O$ defining the output (e.g. $O=\ket{11}\bra{11}$ for 2-qubit system),
Input $x$, trained parameter $\boldsymbol{\theta}$,
Number of shots $N$,

\Ensure
Estimated QNN output $\hat p(x, \boldsymbol{\theta})$

\State Initialize accumulator $s \leftarrow 0$
\For{$j=1, \dots, N$}
    \State Prepare $\ket{0}^{\otimes n}$
    \State Apply $\mathcal{A}(x;\boldsymbol{\theta})$
    \State Measure the qubits required to evaluate $O$
    \State Convert the measurement outcome to a sample $z_j \in \{0,1\}$
    \State $s \leftarrow s + z_j$
\EndFor

\State $\hat p \leftarrow s / N$
\State \Return $\hat p$

\end{algorithmic}
\end{algorithm}

\begin{algorithm}[t]
\caption{Single-shot AE inference for QNN output (AE-QNN)}
\label{alg2}
\begin{algorithmic}[1]

\Require
QNN state-preparation circuit $\mathcal{A}(x;\boldsymbol{\theta})$ acting on $n$ qubits,
Phase flip on ``good'' subspace $S_\chi$,
Phase flip on $\ket{0}^{\otimes n}$ $S_0$,
Evaluation register size $m$,

\Ensure
Estimated QNN output $\hat p(x, \boldsymbol{\theta})$

\State Define Grover iterate $Q \leftarrow -\mathcal{A} S_0 \mathcal{A}^\dagger S_\chi$
\State Prepare $\ket{0}^{\otimes m} \otimes \ket{0}^{\otimes n}$
\State Apply Hadamard $H^{\otimes m}$ on the evaluation register
\State Apply $\mathcal{A}$ on the system register
\For{$k=0, \dots, m-1$}
    \State Apply controlled-$Q^{2^{m-1-k}}$ with control = evaluation qubit $k$
\EndFor
\State Apply inverse QFT on the evaluation register
\State Measure the evaluation register to obtain $z \in \{0, \dots, 2^{m-1}\}$

\State $\hat \phi \leftarrow z/2^m$; $\hat \phi \leftarrow \text{min}(\hat \phi, 1-\hat \phi)$
\State $\hat p \leftarrow \sin^2(\pi \hat \phi)$
\State \Return $\hat p$

\end{algorithmic}
\end{algorithm}

Algorithms \ref{alg1} and \ref{alg2} describe pseudo-codes used in this work, for the MC inference for QNN output (MC-QNN) and AE inference for QNN output (AE-QNN), respectively. These two inference strategies differ primarily in how they trade measurement repetitions against coherent circuit depth, from the same QNN circuit $\mathcal{A}(x; \boldsymbol{\theta})$. MC-QNN uses shallow circuits repeated many times, whereas AE-QNN uses deeper circuits with additional ancilla qubits but can drastically reduce the number of measurement shots required to reach a target accuracy (especially single-shot inference in Algorithm \ref{alg2}). In practice, this trade-off must be evaluated against hardware characteristics: near-term devices with limited coherence may favor MC sampling, while platforms where shots dominate the cost or where repeated sampling is infeasible may benefit from AE-QNN, particularly when inference is performed with a fixed, trained model and high precision is required for a small number of queries.

To enable systematic and interpretable analysis, we consider a minimal yet expressive two-qubit ($n=2$) QNN architecture. This choice allows for exact classical simulation of the noiseless circuit while retaining nontrivial entanglement and parameter dependence.

The QNN (or the oracle in AE) is implemented as a parameterized quantum circuit of the form

\begin{equation}
    \mathcal{A}(x;\boldsymbol{\theta}) = U(\boldsymbol{\theta})U_\text{enc}(x),
\label{eq_oracle} \\
\end{equation}

\noindent acting on an initial state $\ket{0}^{\otimes2}$. The encoding layer 

\begin{equation}
    U_\text{enc}(x)=R_X(\alpha x) \otimes R_Y(\beta x)
\label{eq_oracle} \\
\end{equation}

\noindent with fixed encoding scales $\alpha=1.0$ and $\beta=1.7$ maps the classical input $x$ to continuous rotations of the qubit states and is representative of commonly used data re-uploading strategies in QNNs.

For the trainable unitary $U(\boldsymbol{\theta})=\prod^L_{l=1} U_l(\boldsymbol{\theta}_l)$, each layer $U_l(\boldsymbol{\theta}_l)$ is composed of single-qubit rotations $R_Y(\theta_{l,1}), R_X(\theta_{l,2})$ on each qubit, followed by a fixed entangling gate (CNOT). This yields a total of $4L$ trainable parameters. The circuit structure is expressive enough to represent nonlinear functions of the input while remaining shallow ($L=3$ in this work), which is essential for isolating inference-related effects from circuit-depth limitations.

The QNN output is defined as the probability of projecting the final two-qubit state onto the computational basis state $\ket{11}$ (or equivalently $O=\ket{11}\bra{11}$). This binary projector output allows direct comparison between Monte Carlo sampling and amplitude estimation, and it provides a clear interpretation of estimation error in terms of probability deviation.

\subsection{Resource advantage}\label{sec3.3}

In this section, we quantitatively compare the inference accuracy of MC-QNN and AE-QNN under controlled resource budgets. To isolate inference-related effects from training dynamics, we consider a frozen QNN whose parameters are randomly initialized and fixed throughout the experiments. This setting allow us to focus exclusively on the efficiency of the readout algorithms. Because the experiments are performed using quantum circuit simulation, the exact expectation value of the QNN output is known. This enables a direct and unambiguous evaluation of the inference error for both MC-QNN and AE-QNN, defined as the absolute deviation between the estimated output and the exact value.

\begin{figure}[t!]
\centering
\includegraphics[width=\linewidth]{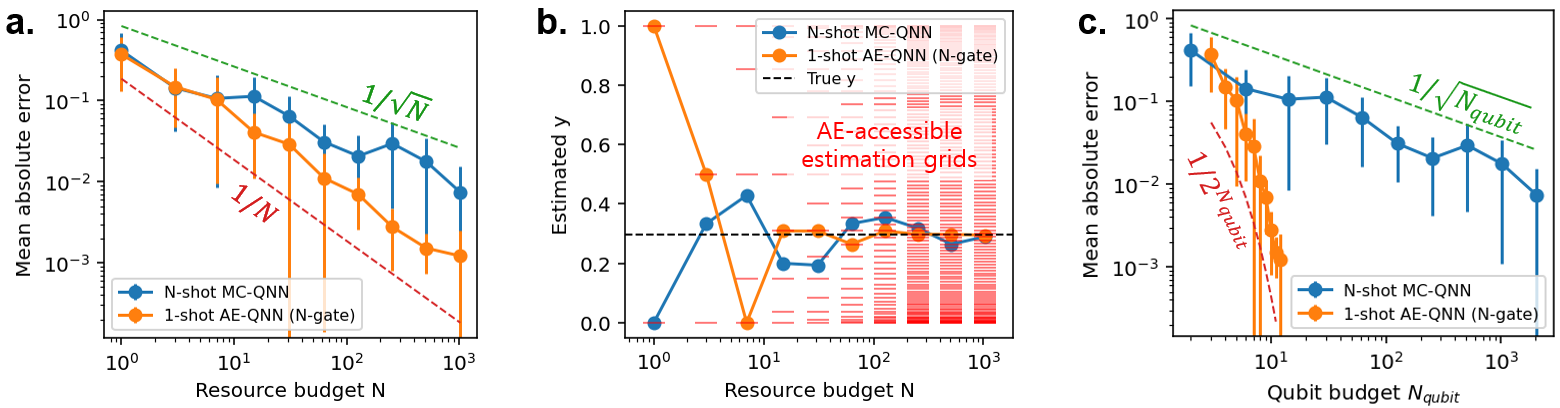}
\caption{Comparison of the inference accuracy of the MC-QNN and AE-QNN. \textbf{a}. Inference error as a function of the computational resource budget $N$. \textbf{b}. Example estimation of MC-QNN and AE-QNN for a given target, with respect to the resource budget $N$. \textbf{c}. Inference error as a function of the qubit budget $N_\text{qubit}$. }\label{fig2}
\end{figure}

Figure \ref{fig2}a compares the inference accuracy of MC-QNN and AE-QNN as a function of the total computational resource budget $N$. For MC-QNN, this resource budget $N$ corresponds to the number of measurement shots, whereas for AE-QNN it corresponds to the number of controlled Grover-iterate calls, which dominate the computational cost of the amplitude estimation circuit. For each value of $N$, the inference error is averaged over ten randomly chosen input values, and the corresponding standard deviation is also marked in the figure.

As expected from statistical sampling theory, the error of MC-QNN (blue) decreases as $\mathcal{O}(1/\sqrt{N})$. This behavior reflects the binomial nature of shot-based sampling and is consistent with the standard Monte Carlo estimator for expectation values. In contrast, AE-QNN (orange) exhibits a markedly faster error reduction, following $\mathcal{O}(1/N)$ with respect to the query count. This quadratic improvement in query efficiency is consistent with the theoretical guarantees of quantum amplitude estimation. Notably, AE-QNN already saturates the theoretical limit even in the single-shot regime, and increasing the number of measurement shots does not lead to a dramatic reduction in estimation error. This behavior arises because the coherent Grover iterations effectively amplify the relevant probability amplitude prior to measurement, leading to a saturation of estimation accuracy once sufficient amplification has been achieved.

To further analyze the qualitative difference between the two inference strategies, Figure \ref{fig2}b shows the convergence behavior for a single representative input as the query budget increases. For MC-QNN, increasing the number of shots leads to progressively finer sampling of the underlying Bernoulli distribution. The estimated output converges smoothly toward the exact value as dictated by the law of large numbers, with fluctuations diminishing at a rate proportional to $1/\sqrt{N}$. In contrast, AE-QNN exhibits a distinct convergence pattern. As the query budget increases, the effective number of qubits $m$ in the evaluation register increases, thereby enlarging the number of discretization grids $2^m$ (red bars in Figure \ref{fig2}b) available for representing the estimated phase. Each additional evaluation qubit refines the resolution of the amplitude estimate, allowing the algorithm to select an approximation that lies closer to the true value. Consequently, AE-QNN converges toward the correct output by successively refining the discretized estimate rather than by averaging over many stochastic samples. In particular, due to the characteristics of the phase estimation and the sinusoidal mapping in Equation \ref{eq_amplitude_recover}, the estimation grids are more densely allocated near 0 and 1, as seen in Figure \ref{fig2}b. This suggests that AE-QNN can be more advantageous than MC-QNN in scenarios involving rare-event detection or the accurate prediction of very small probabilities.

This difference highlights a fundamental contrast between the two approaches. MC-QNN relies on repeated stochastic measurements to suppress noise, whereas AE-QNN trades measurement repetition for coherent circuit depth, extracting information through quantum interference. As a result, our AE-QNN can achieve high-precision inference with dramatically fewer measurement shots, provided that coherent operations can be executed reliably.

The observed scaling behavior demonstrates a clear resource advantage of AE-QNN in inference tasks where measurement shots constitute a dominant cost. In particular, the ability of AE-QNN to reach near-saturated accuracy in the few-shot or single-shot regime is especially relevant for platforms such as photonic quantum devices, where each shot requires the physical generation of new qubits. In such settings, the quadratic reduction in effective query complexity (or, reduction from $\mathcal{O}(N)$ to $\mathcal{O}(1)$ in terms of the number of measurement shots) can translate directly into substantial experimental savings. Figure \ref{fig2}c shows the inference accuracy of MC-QNN and AE-QNN with respect to the number of required qubit generations $N_\text{qubit}$. In terms of qubit generation, AE-QNN with single-shot inference (error of $\mathcal{O}(1/2^{N_\text{qubit}})$) can provide much higher accuracy than MC-QNN (error of $\mathcal{O}(1/\sqrt{N_\text{qubit}})$) when the number of qubit generations is limited.

At the same time, the deeper circuits required by AE-QNN imply increased sensitivity to gate-level noise and coherence limitations. Therefore, the practical advantage of AE-QNN depends on the balance between shot cost and circuit fidelity, a trade-off that we examine further in the subsequent sections.

\subsection{Effect of the quantum error on the inference}\label{sec3.4}

Beyond finite-shot sampling noise, near-term quantum hardwares suffer from several sources of intrinsic noise, such as decoherence \cite{EtxezarretaMartinez2021_decoherence, PhysRevX.13.041022_decoherence, Wang2024canerrormitigation_dephasing}, gate operation imperfections \cite{Dalton2024_gate_error, PhysRevA.110.012404_gate_error}, and readout errors \cite{doi:10.1126/sciadv.abi8009_readout_error, PhysRevA.107.062426_readout_error, Cheng2023_nisq}. Unlike sampling noise, these hardware-induced errors accumulate with circuit depth and cannot be mitigated by increasing the number of measurements, even for AE-QNNs.

To emulate near-term quantum hardware with these intrinsic errors, we include a depolarizing noise model applied after each elementary operation. Specifically, after every single-qubit and two-qubit gate, the system undergoes a depolarizing channel

\begin{equation}
    \rho \leftarrow (1-p_\text{error})\rho + \frac{p_\text{error}}{3} \left( X\rho X^\dagger + Y\rho Y^\dagger + Z\rho Z^\dagger \right),
\label{eq_deph_noise} \\
\end{equation}

\noindent where $\rho$ is the qubit density matrix at each layer, $p_\text{error}$ denotes the error rate and $n$ is the number of qubits acted upon. This noise channel randomly depolarizes the qubit with respect to Pauli-X, Y, or Z gates, with probability $p_\text{error}$.

This noise model captures generic gate imperfections while remaining hardware-agnostic. Measurement noise is implicitly included through repeated sampling or AE readout repetition, allowing us to study the combined effect of gate-level noise and inference strategy.

\begin{figure}[t!]
\centering
\includegraphics[width=\linewidth]{fig3}
\caption{Comparison of inference error of MC-QNN and AE-QNN when the quantum device has a gate error. \textbf{a}. Inference error as a function of $p_\text{error}$. \textbf{b}. Inference error with respect to the computational resource budget $N$, when $p_\text{error}=0.01$.}\label{fig3}
\end{figure}

Figure \ref{fig3}a compares the inference accuracy of MC-QNN and AE-QNN in the presence of gate-level noise. The experimental setup is identical to that of Section \ref{sec3.3}, except that depolarizing errors are inserted after each gate operation to model imperfect quantum hardware. The horizontal axis represents the per-gate error rate, while the vertical axis shows the absolute inference error with respect to the exact expectation value obtained from noiseless simulation.

The final prediction error in both inference schemes arises from the accumulation of gate-level errors along the quantum circuit. However, the manner in which these errors accumulate differs substantially between MC-QNN and AE-QNN due to their distinct circuit structures. In MC-QNN, inference consists of repeatedly executing a relatively shallow QNN circuit and averaging the measurement outcomes. As a result, increasing the gate error rate leads to a gradual, approximately linear degradation in accuracy. Each circuit execution incurs a similar amount of noise, but the overall effect remains moderate because the circuit depth is fixed and relatively small.

In contrast, AE-QNN relies on coherent amplitude amplification implemented through repeated Grover iterations. Each Grover iteration itself contains multiple applications of the QNN unitary and its inverse, as well as additional reflection operations. Consequently, the total number of gates traversed during a single AE-QNN inference is significantly larger than that of MC-QNN. As the gate error rate increases above a certain threshold $\sim 10^{-2}$, the cumulative effect of these errors grows rapidly, leading to a much stronger sensitivity of the final prediction accuracy to noise. This behavior is clearly visible in Figure \ref{fig3}a, where the AE-QNN error increases more steeply with the error rate than that of MC-QNN when $p_\text{error}>10^{-2}$.

An additional factor influencing the noise sensitivity of AE-QNN is the depth of the evaluation register. In Section \ref{sec3.3}, increasing the number of evaluation qubits $m$ could improve the resolution of the estimation grids. However, a deeper evaluation register also requires a greater number of controlled Grover operations, each of which introduces further opportunities for gate errors to accumulate. As a result, when the gate error rate becomes non-negligible, AE-QNN with a large number of evaluation qubits exhibits a more pronounced degradation in performance. This trade-off is evident in Figure \ref{fig3}b, which compares the inferences by MC-QNN and AE-QNN when the per-gate error rate is $0.01$. Both MC-QNN and AE-QNN follow the theoretical scaling of resource $N$ when $N$ is moderate, where the accumulated gate error is not a dominant issue. However, the inference error of AE-QNN increases rapidly when $N \gtrsim 10^2$ (or $m\gtrsim7$), deviating the theoretical scaling, $\mathcal{O}(1/N)$.

This noise sensitivity constitutes a clear drawback of AE-QNN relative to MC-QNN and is consistent with a well-known limitation of Grover-based quantum algorithms in general. Similar behavior has been extensively studied in the context of Grover search \cite{Ijaz2023_grover_noise}, where repeated coherent iterations amplify not only the desired signal but also the impact of noise. Accordingly, achieving reliable performance with AE-QNN requires sufficiently low gate error rates, making improvements in hardware fidelity particularly important for practical deployment. Nevertheless, there are substantial research efforts devoted to mitigating noise in Grover-type algorithms, such as variants of Grover iterations \cite{Grinko2021_iqae}, error-aware or noise-resilient amplitude amplification \cite{Tanaka2021_noisy_qae}, and error mitigations \cite{RevModPhys.95.045005_error_mitigation}. These approaches aim to preserve the favorable scaling of AE while reducing its vulnerability to accumulated gate errors.

We also note that despite its increased noise sensitivity, Figure \ref{fig3} demonstrates that AE-QNN remains more accurate than MC-QNN in the low-noise ($p_\text{error}\lesssim 10^{-2}$) and moderate resource ($N \lesssim 10^2$) regime relevant to near-term devices. In current near-term quantum hardware, typical per-gate error rates range from $10^{-3}$ to $10^{-2}$ \cite{PRXQuantum.3.040326_quantum_error_rate}, with two-qubit gates generally exhibiting higher error rates than single-qubit operations. Therefore, within realistic near-term conditions, AE-QNN can still offer a competitive inference accuracy, particularly in scenarios where measurement shots are costly or severely limited.

\subsection{Single-shot training}\label{sec3.5}

The previous sections have focused on ``inference'' using a trained QNN. In that setting, AE-QNN demonstrated that single-shot readout can already achieve an accuracy comparable to, or exceeding, that of MC-QNN requiring tens of thousands of measurement shots. However, training of QNN presents a fundamentally different situation. Unlike classical neural networks, which rely on automatic differentiation and backpropagation through a fully differentiable computational graph, QNN training typically employs the parameter-shift rule \cite{Mitarai_PRA2018_param_shift, Schuld_PRA2019_param_shift} to compute gradients with respect to circuit parameters. For each trainable parameter, the circuit must be evaluated at two shifted parameter values ($\pm$), and the gradient is obtained from the difference of the corresponding expectation values. As a result, a single update step of full parameters inevitably requires multiple circuit evaluations proportional to the number of parameters, regardless of the inference strategy used for each evaluation.

Furthermore, during training, sampling noise is not purely detrimental. In many cases, stochasticity in gradient estimates acts as an implicit regularizer, stabilizing optimization and preventing overfitting. Consequently, even in MC-QNN training, it is often unnecessary to use a large number of shots per parameter shift. This reduces the practical benefit of AE-QNN relative to inference, where shot efficiency directly corresponds to improved accuracy. As a result, the advantages of AE-QNN are less pronounced during training than during inference.

Nevertheless, from the perspective of the total number of required shots, AE-QNN can still offer a substantial reduction compared to MC-QNN. In MC-QNN, each parameter-shift evaluation still requires $N\gtrsim \mathcal{O}(10)$ measurement shots to estimate an expectation value with acceptable precision. In contrast, AE-QNN can replace these $N$ shots with a single-shot readout, albeit at the cost of additional Grover iterations within the circuit. This distinction becomes particularly important on platforms where the cost of each shot is high. On photonic quantum platforms, conventional MC-QNN training is effectively infeasible due to the prohibitive cost of repeated qubit generation, whereas AE-QNN training may offer a viable alternative. This section aims to assess whether training with AE-QNN is practically feasible under realistic resource constraints.

\begin{figure}[t!]
\centering
\includegraphics[width=4in]{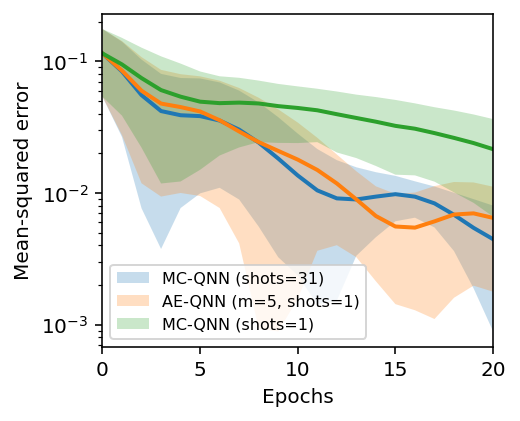}
\caption{The loss history during training the MC-QNN (blue) and AE-QNN (orange) with the same computational resource. For comparison, the single-shot training of MC-QNN (green) is also shown.}\label{fig4}
\end{figure}

Figure \ref{fig4} compares the training loss curves obtained using MC-QNN and AE-QNN under the same resource budget per epoch ($N=2^5-1=31$). For MC-QNN, each parameter shift uses $N$ measurement shots. For AE-QNN, each parameter shift uses a single-shot readout, while the corresponding evaluation requires $N$ Grover iterations implemented coherently within the circuit. Despite this stark difference in how resources are allocated, AE-QNN-based training does not underperform MC-QNN-based training. Even in the single-shot regime, AE-QNN achieves comparable convergence behavior, demonstrating that training remains feasible when shot-based sampling is replaced by coherent AE. This suggests that AE-QNN can enable QNN training on platforms where multi-shot sampling is prohibitively expensive.

An interesting observation is that even MC-QNN training with single-shot inference (green) where each forward evaluation yields only a binary outcome can still make progress toward minimizing the loss. Although convergence is significantly slower than in the single-shot AE-QNN case, the fact that learning remains possible highlights the robustness of variational optimization in QNNs. A possible explanation is that, when aggregated over many parameter updates, the stochastic gradients obtained from binary outcomes still encode directional information about the loss landscape. A detailed theoretical analysis of this phenomenon is beyond the scope of the present work and is left for future investigation.


However, it is important to emphasize a key trade-off. In MC-QNN, multiple measurement shots can often be parallelized, especially in multi-qubit systems or experimental setups that support batch execution. As a result, increasing the number of shots does not necessarily induce a proportional increase in wall-clock training time. In contrast, AE-QNN relies on serial Grover iterations, whose number grows with the desired precision. This serial structure directly impacts circuit depth and execution time, making AE-QNN training slower in terms of coherent quantum evolution. In this sense, AE-QNN reduces the number of shots at the expense of increased circuit depth and runtime.

\section{Summary and discussion}\label{sec4}

In this work, we investigated the fundamental resource costs associated with inference and training in quantum neural networks and demonstrated how these costs can be substantially reduced by integrating quantum amplitude estimation into the QNN readout stage. Conventional QNN inference relies on Monte Carlo sampling of measurement outcomes, leading to an intrinsic $\mathcal{O}(1/\sqrt{N})$ sampling error and necessitating a large number of repeated circuit executions. This requirement poses a significant obstacle for near-term quantum hardware and becomes particularly prohibitive on photonic platforms, where each shot entails costly qubit generation.

By embedding a frozen QNN within an amplitude estimation protocol, we introduced an alternative inference strategy (AE-QNN) that replaces repeated stochastic sampling with coherent quantum interference. Through systematic numerical experiments, we showed that AE-QNN achieves a quadratic improvement in query efficiency, exhibiting an $\mathcal{O}(1/N)$ error scaling with respect to the number of Grover queries. Remarkably, this advantage manifests already in the single-shot regime, where AE-QNN attains accuracy comparable to or exceeding that of MC-QNN using tens to thousands of shots. The discrete grid structure inherent to phase estimation further endows AE-QNN with favorable resolution near small and large probabilities, making it especially suitable for rare-event detection and low-probability inference tasks.

We also examined the impact of intrinsic quantum noise on both inference strategies. While AE-QNN is more sensitive to gate-level errors due to its increased circuit depth and repeated Grover iterations, we found that its performance remains competitive within realistic near-term error rates ($10^{-3}-10^{-2}$ per gate). These results clarify the trade-off between shot cost and circuit fidelity, and delineate the operational regime in which AE-QNN provides a net advantage.

Beyond inference, we explored the feasibility of training QNNs in the single-shot regime. Although training inherently requires multiple circuit evaluations per update due to the parameter-shift rule, AE-QNN can still reduce the total number of measurement shots by more than an order of magnitude compared to MC-QNN. We demonstrated that AE-QNN-based training achieves comparable convergence under equal resource budgets, thereby enabling QNN training on platforms where multi-shot sampling is impractical.

Overall, this study shows that the apparent sampling overhead of quantum neural networks is not a fundamental limitation but rather a consequence of conventional readout strategies. By leveraging quantum amplitude estimation, both inference and training costs can be significantly reduced, opening new possibilities for practical quantum machine learning on near-term and shot-limited quantum hardware. Our findings underscore the importance of algorithmic co-design between quantum models and quantum algorithms, and point toward a broader paradigm in which quantum resources are optimally allocated across state preparation, evolution, and readout.

\section*{Author contributions}
J.S. contributed to the design of this study, numerical experiments, analyses, and writing the manuscript.

\section*{Data availability}
The codes that generate data in this paper will be available at the GitHub repository, https://github.com/jaem-seo/single-shot-qnn.

\section*{Declaration of competing interest}
The authors declare that they have no known competing financial interests or personal relationships that could have appeared to influence the work reported in this paper.

\section*{Acknowledgments}
This work was supported by a National Research Foundation of Korea (NRF) grant funded by the Korea government (MSIT) (Grant No. RS-2024-00346024).




\bibliographystyle{elsarticle-num} 
\bibliography{mybib.bib}

\end{document}